\documentclass[10pt,emptycopyrightspace]{ewsn-proc}

\usepackage{xspace}
\usepackage{balance}
\usepackage{graphicx}
\usepackage[hyphens]{url}
\usepackage{epstopdf}
\usepackage{microtype}
\usepackage{booktabs}
\usepackage[table,xcdraw]{xcolor}
\usepackage{caption}
\usepackage{subcaption}
\usepackage{amsmath}
\usepackage{enumitem}
\usepackage{amssymb}
\usepackage[numbers]{natbib}
\usepackage{tikz}
\usepackage{pifont}
\usepackage{graphics}
\usepackage{array}
\usepackage{algorithm,algorithmic,refcount}
\usepackage{marginnote}
\usepackage[hang,flushmargin]{footmisc}
\usepackage{supertabular}

\newcolumntype{P}[1]{>{\centering\arraybackslash}p{#1}}
\newcolumntype{M}[1]{>{\centering\arraybackslash}m{#1}}

\newcommand{\fakepar}[1]{\smallbreak\noindent{}}
\newcommand{\boldpar}[1]{\smallbreak\noindent\textbf{#1.}}

\newcommand{\wifi}{\mbox{Wi-Fi}\xspace}
\newcommand{\aware}{\mbox{Wi-Fi~Aware}\xspace}
\newcommand{\ftm}{\mbox{Wi-Fi~FTM}\xspace}
\newcommand{\rtt}{\mbox{Wi-Fi~RTT}\xspace}
\newcommand{\eleven}{\mbox{IEEE\,802.11}\xspace}

\usepackage{manyfoot}
\DeclareNewFootnote{A}
\DeclareNewFootnote{B}

\let\footnoteR\footnoteB
\let\footnote\footnoteA
\iftrue
    \newcommand{\comment}[1]{\footnoteR{{\color{red} #1}\color{red}}}
\else
    \newcommand{\comment}[1]{}
\fi

\begin{document}

% Title
\title{
Benchmarking and Security Considerations \\ of Wi-Fi FTM for Ranging in IoT Devices
}

% Authors (blinded)
% \numberofauthors{5}
\iffalse
\author{
\alignauthor Double Blind \\
  \affaddr{do not reveal authors}
}
\else
\author{
\alignauthorpage Govind Singh$^{*}$, Anshul Pandey$^{*}$, Monika Prakash$^{*}$, Martin Andreoni$^{*}$, and Michael Baddeley$^{*}$ \\
    \affaddr{$^{*}$Technology Innovation Institute, UAE}\\
    \email{\{govind.singh; anshul.pandey; monika.prakash; martin.andreoni; michael.baddeley\}@tii.ae}
}
\fi

% make the title area
\maketitle

\begin{abstract}
% Since the introduction of IEEE\,802.11mc an increasing number of commercial \wifi vendors employ the standard for device localization. Specifically, 
\noindent The IEEE\,802.11mc standard introduces fine time measurement (\ftm), allowing high-precision synchronization between peers and round-trip time calculation (\rtt) for location estimation -- typically with a precision of one to two meters. This has considerable advantages over received signal strength (RSS)-based trilateration, which is prone to errors due to multipath reflections. We examine different commercial radios which support \rtt and benchmark \ftm ranging over different spectrums and bandwidths. Importantly, we find that while \ftm supports localization accuracy to within one to two meters in ideal conditions during outdoor line-of-sight experiments, for indoor environments at short ranges similar accuracy was only achievable on chipsets supporting \ftm on wider (VHT80) channel bandwidths rather than narrower (HT20) channel bandwidths. Finally, we explore the security implications of \ftm and use an on-air sniffer to demonstrate that FTM messages are unprotected. We consequently propose a threat model with possible mitigations and directions for further research.
\end{abstract}


%----------------------------------------------%
\section{Introduction}
\label{sec:introduction}
%----------------------------------------------%

\noindent The fine time measurement (\ftm) and round-trip-time \rtt features of the IEEE\,802.11mc amendment to the \wifi standard allow supporting devices to estimate the distance to other devices supporting \ftm ranging. This capability is supported in various modes (\textit{Station (STA)}, \textit{Access Point (AP)}, and \textit{Neighborhood Aware Networking (NAN)}) and allows applications to benefit from accurate localization and context awareness in use-cases such as asset tracking, geofencing, industrial robotics, home automation control, and location-based information for service broadcasting.

Traditionally, distance ranging based on the Received~Signal~Strength~Indicator~(RSSI), which estimates the power of the received signal as measured through the hardware, is a simple and extensively adopted means for estimating distance -- often employing a straightforward long-distance path loss model~\cite{zafari2019survey},

\vspace{-5mm}
\begin{equation}
    \mbox{RSSI} = -10*n\log10(d) + A
    \label{eq:equation_1}
\end{equation}

\begin{table*}[ht]
    \caption{Parameters affecting \ftm ranging performance.}
    \label{tab:ftm_parameters}
    \footnotesize
    \renewcommand{\arraystretch}{1.1}
    \centering
    \begin{tabular}{ c l }
        \toprule
        \textbf{FTM parameters} & \textbf{Description} \\
        \midrule
        Frequency band    & 6GHz/5GHz has better accuracy in comparison to 2.4GHz due to available bandwidth.   \\
        Channel bandwidth & Ranging error roughly halves with double bandwidth, when comparing the 20/40/80/160 MHz frequency bands. \\
        Antenna diversity & Multiple antennas reduce ranging error due to better diversity. 2x2 configuration has approx 30 percent better accuracy. \\
        Multipath         & Line-of-sight path is preferred for the best accuracy. \\
        Tx power          & Higher transmission power allows for minimizing the errors when used within the regulatory limit. \\
        Timestamping      & Hardware time stamping clock resolution and lower ppm errors allow better accuracy.   \\
    \end{tabular}
    \vspace{-4.00mm}
\end{table*}

where $A$ is the RSSI at the reference distance and $n$ is a path loss exponent. While this method is easy to employ, due to multipath issues the accuracy is frequently poor as the relationship between RSSI and distance is non-linear and is adversely affected by obstacles in non line-of-sight environments.
In contrast, precise time measurement provided by \ftm, \rtt calculates the distance between two devices by measuring the time a packet takes to make a round trip (i.e., the time taken for an initiating device to receive an acknowledgment from a neighbor after transmitting a packet to that neighbor). The difference between the transmit and receive time stamps denotes the flight time which, when taking into account the speed of light, indicates how far away the \rtt initiator device is from the \rtt responder device. Importantly, RTT measurement errors are linear with distance, not exponential like RSSI, allowing sub-meter accuracy. However, precise time-stamping of packets supported by \ftm is highly dependent on the hardware capabilities, such as the clock resolution. Additionally, path First Arrival Correction (FAC) and accurate time-stamping of the first packet symbol are also key contributors to \rtt accuracy. The aim of this work is to therefore evaluate the capabilities of different commercial off-the-shelf (COTS) hardware platforms supporting the \ftm protocol and comparing the ranging accuracy across different bandwidth configurations.

\boldpar{Our contributions} In this paper we identify a number of commercially available \wifi chipsets which support \ftm ranging and benchmark the accuracy in both outdoor and indoor experimental setups.

\boldpar{Outline of this paper} In Section~\ref{sec:background} we provide necessary preliminaries on the operation of \rtt and \ftm. In Section~\ref{sec:experimental_analysis} we provide details of our experimental setup and perform benchmarking of three commercially available \ftm chipsets. We explore related localization technologies in Section~\ref{sec:related_work} before concluding in Section~\ref{sec:conclusions}.
%----------------------------------------------%
\section{Ranging with \ftm}
\label{sec:background}
%----------------------------------------------%

\begin{figure}[t]
	\centering
        \begin{subfigure}[t]{.75\columnwidth}
            \centering
    	\includegraphics[width=.8\columnwidth]{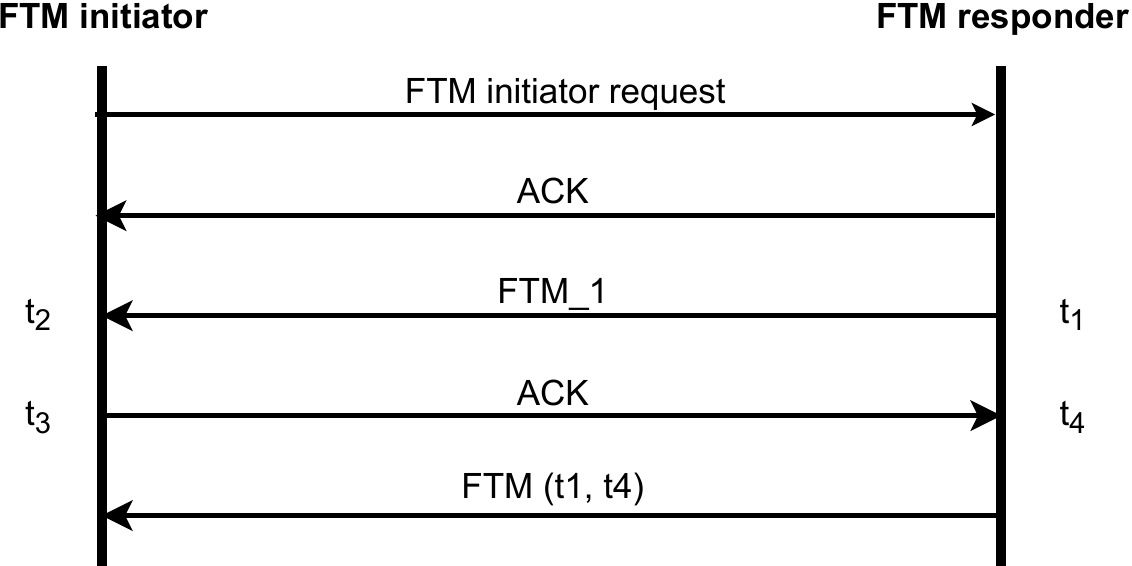}
    	\caption{Single mode RTT protocol.}
            \vspace{2.00mm}
    	\label{fig:ftm_single}
        \end{subfigure} 
        \begin{subfigure}[t]{.8\columnwidth}
            \centering
    	\includegraphics[width=.75\columnwidth]{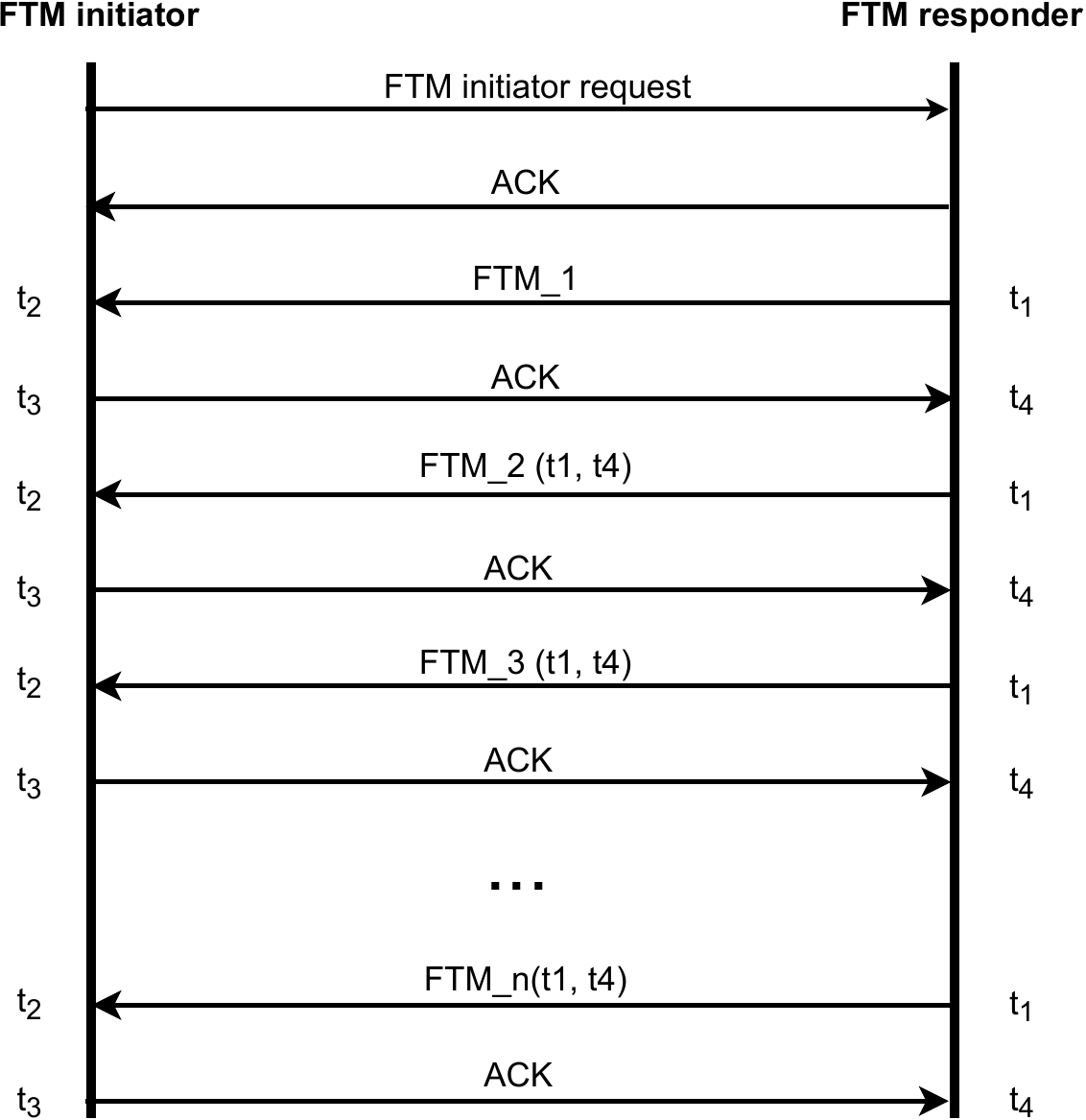}
    	\caption{Burst mode RTT protocol.}
    	\label{fig:ftm_burst}
        \end{subfigure}
        \caption{\ftm protocol in \textit{single} and \textit{burst} modes.}
        \vspace{-6.00mm}
        \label{fig:ftm_single_burst}
\end{figure}

\noindent \ftm was introduced in the IEEE\,802.11-2016 standard revision for the purpose of allowing a fine timing measurement (FTM) frame to calculate round-trip time (RTT) between two peers. Specifically, \rtt makes use of \eleven frame exchange based on acknowledgment frames (as per Figure~\ref{fig:ftm_single}) as well as FTM bursts (as per Figure~\ref{fig:ftm_burst}), which further allows delivering many FTM messages consecutively while lowering timing computation error by averaging multiple timestamps.

In both instances, an FTM request frame (\texttt{FTMR}) is provided by the FTM \textit{initiator} to start the process, while the FTM responder has the option of accepting it or rejecting the \texttt{FTMR}. Upon acceptance, the \textit{responder} sends an acknowledge (\texttt{ACK}) frame. The \textit{responder} then sends a single \texttt{FTM} frame, recording the current time $t_1$ within the frame, to start the beginning of the round-trip measurement. As soon as the preamble is detected by the initiator, it starts recording time at time $t_2$. The FTM initiator then creates its own \texttt{ACK} frame and transmits it to the responder. The \texttt{ACK} frame is transmitted at the initiator, which accounts for the hardware signal processing delay of the initiator's digital baseband and radio front end. Furthermore, the responder can determine the overall round-trip time of the signal, as per Equation~\ref{eq:rtt}, by deducting $t_1$ from $t_4$ when it receives the initiator's \texttt{ACK} frame at time $t_4$ and adding this to the delta between $t_3$ and $t_2$.

\vspace{-3.00mm}
\begin{equation}
    \mbox{RTT} = (t4-t1 + t2-t3)
    \label{eq:rtt}
\end{equation}

The time a WiFi signal takes to travel over the air between the RTT initiator peer and the RTT responder peer is proportional to the actual distance between them (approximately 3.3 nanoseconds per meter). Accordingly, the distance between the peers can be estimated using Equation~\ref{eq:distance}.

\vspace{-3.00mm}
\begin{equation}
    d =\mbox{ RTT}/2*c = (t4-t1 + t2-t3)/2 * c
    \label{eq:distance}
\end{equation}

% Each responder peer in a Wi-Fi Location network is configured with its exact location, including geospatial coordinates (latitude, longitude, and altitude) and civic address. This allows more precise location determination than with other solutions, even in multilevel structures.

\begin{table*}[!t]
    \caption{Experimental Setup Configurations.}
    \label{tab:setup_configurations}
    \footnotesize
    \renewcommand{\arraystretch}{1.0}
    \centering
    \begin{tabular}{ c c c c c c c}
        \toprule
        Configuration & Initiator                  & Responder                      & Specification & Channel  & Bursts     & FTM Ranging \\
        \midrule
        Config. 1     & Google Pixel 4a (WCN3990)  & Google Wi-Fi Mesh AP (QCA4019) & VHT80         & 5745MHz  & 8          &  Native     \\
        Config. 2     & FeatherS2 NEO (ESP32-S2)   & FeatherS2 NEO (ESP32-S2)       & HT20          & 2412MHz  & 2          &  Native     \\
        Config. 3     & Google Pixel 4a (WCN3990)  & Google Pixel 4a (WCN3990)      & HT20          & 2412MHz  & 8          & \aware      \\
    \end{tabular}
    \vspace{-4.00mm}
\end{table*}

However, in some circumstances (e.g., due to environmental factors such as temperature~\cite{boano2014templab}) the internal clock between the initiator peer with the responder peer may not be synchronized, hence direct subtraction can not be done with two timestamps to calculate RTT. The difference in timestamps when the signal travels in the reverse direction is affected in the opposite way by the clock offset between peers. As a result, the round trip time (RTT) can be obtained without having to know the clock offsets, by simple addition and subtraction of four timestamps (as per Equation~\ref{eq:distance}).

Fundamentally, however, localization accuracy depends not only on the accuracy of the timestamp but also on numerous other factors. The bandwidth, antenna diversity, the transmit power of the broadcasting device, and the reception sensitivity of the receiving device are also key factors of a Wi-Fi radio that define the accuracy of its individual measurements in an RF ranging system. 

\boldpar{Bandwidth} In a wireless system, preamble detection is used to determine the beginning of a frame. Hardware timestamping systems based on preamble detection present a resolution bound equal to the baseband sampling period, on which the FTM protocol depends. Indeed, a detection delay of as little as 1~ns could result in an error of 30 cm (\ref{eq:distance}). Hence, high resolution of the time-stamping clock is required as FTM requires pico second clock resolution for time-stamping. As the sampling period is inverse of the bandwidth (either 20MHz, 40MHz, or 80MHz), better resolution can be achieved at higher bandwidths (as per the Nyquist rate $f_s = 2B$, where $f_s$ is the sampling frequency and $B$ is the bandwidth in MHz).

\boldpar{Antenna diversity} Having multiple antennas helps to extract the spatial diversity (transmitting multiple copies of the same information via multiple antennas) which provides more resilient transmissions, and also helps to save the system from undergoing deep fade events. 

\boldpar{Multipath} Furthermore, the channel frequency response of the associated channel varies owing to the multipath propagation. Specifically, multiple copies of the same signal with different amplitudes and phases arrive at the receiver due to the interaction of the signal with the surrounding objects. Also, due to the multipath propagation, the antenna receives several copies of the transmitted signals that have been subjected to various delays and attenuation. In the frame detector, these reflections can cause a sizable amount of jitter, especially in non-line-of-sight circumstances. FTM accuracy relies on the first arrival correction algorithm due to the multipath behavior, which is generally implemented in the lower MAC of the firmware component.

\boldpar{Timestamping} The coherence time, which specifies the period of time during which the channel is thought to be invariant, describes the dynamic of the channel. This channel variation may cause inaccuracy in the estimation of the channel delay due to the non-stationary state and, as a result, it will affect the FTM accuracy. With mobility, the accuracy can further suffer as the involved nodes' mobility may incur more estimation errors. Finally, with chipsets such as the \texttt{WCN3990} in the Google Pixel 4a including \ftm ranging functionality as part of the \aware specification, there has been little study on \ftm performance when being used as part of a wider standard rather than a `native' approach where the FTM functions are accessed directly.
%----------------------------------------------%
\section{Experimental Analysis}
\label{sec:experimental_analysis}

% \begin{table}[t]
%     \caption{FTM-enabled chipsets.}
%     \label{tab:ftm_chipsets}
%     \footnotesize
%     \renewcommand{\arraystretch}{.9}
%     \centering
%     \begin{tabular}{ c c c c }
%         \toprule
%         chipset           & Vendor    & Mode & Specification \\
%         \midrule
%         ESP32-S2 2.4GHz   & Espressif & I/R  & HT20  \\
%         WCN3990  2.4/5GHz & Qualcomm  & I    & VHT80 \\
%         WCN3980  2.4/5GHz & Qualcomm  & I    & VHT80 \\
%         QCA4019  2.4/5GHz & Qualcomm  & R    & VHT80  \\
%         AC2200   2.4/5GHz & Intel     & I    & VHT80 \\
%     \end{tabular}
% \end{table}

\begin{figure}[t]
        \centering
        \begin{subfigure}[t]{.7\columnwidth}
            \includegraphics[width=1.\columnwidth]{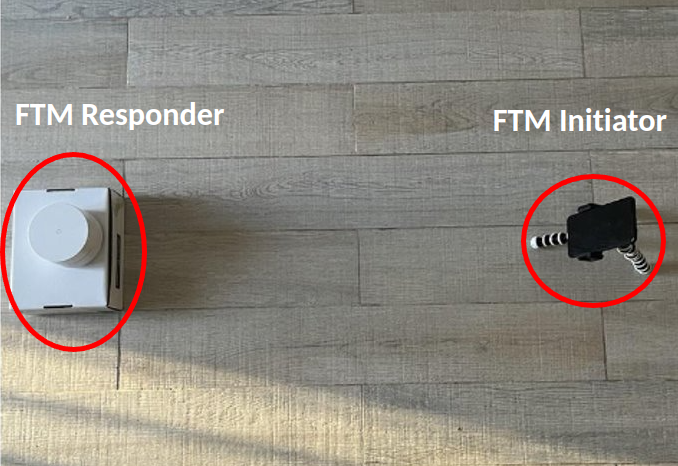}
            \caption{Indoor setup.}
            \vspace{4.00mm}
            \label{fig:indoor_setup}
        \end{subfigure} 
        \begin{subfigure}[t]{.7\columnwidth}
            \includegraphics[width=1.\columnwidth]{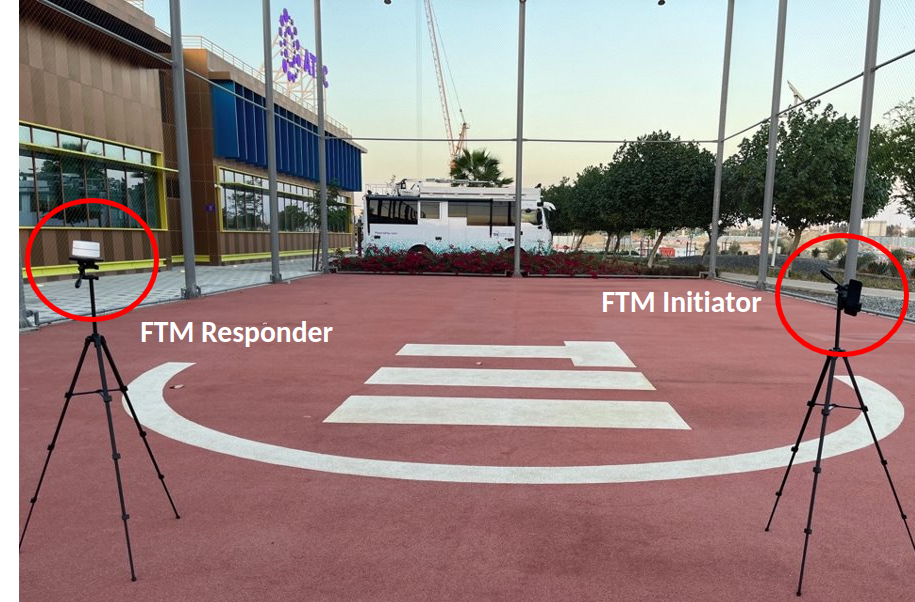}
            \caption{Outdoor setup.}
            \vspace{4.00mm}
            \label{fig:outdoor_setup}
        \end{subfigure} 
        \begin{subfigure}[t]{1.\columnwidth}
            \includegraphics[width=1.\columnwidth]{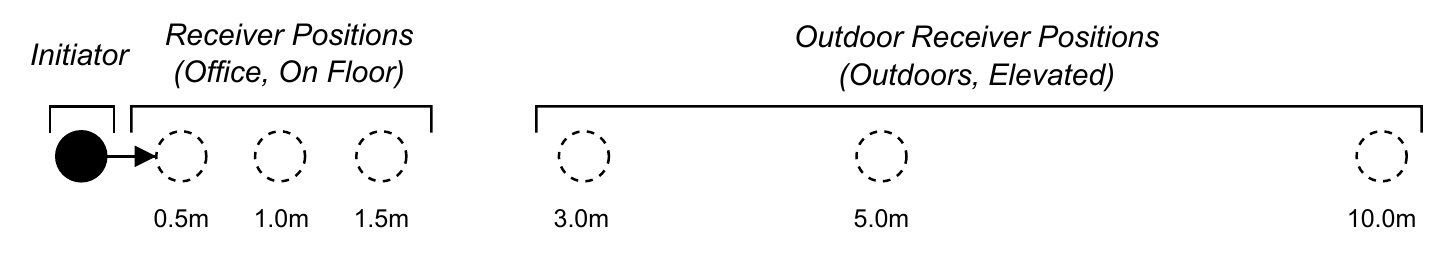}
            \caption{Distances between \ftm \textit{initiator} and \textit{responder}.}
            % \vspace{1.00mm}
            \label{fig:setup_distances}
        \end{subfigure} 
        \caption{Experimental setup.}
        \vspace{-4.00mm}
        \label{fig:indoor_outdoor_setup}
\end{figure}

\noindent To date, there are few examples of commercial \wifi chipsets which support \ftm. In Table~\ref{tab:setup_configurations} we identify three compliant chipsets and indicate the supported channel bandwidth specification, the channel frequency used in experiments, the number of \ftm bursts supported, and whether \ftm ranging is directly accessible in the firmware or is included as part of \aware. 

Specifically, we consider the \texttt{WCN3990} on a Google Pixel 4a, the \texttt{QCA4019} on a Google \wifi Mesh AP, and the \texttt{ESP32-S2} on the FeatherS2 NEO and benchmark these devices in both an \textit{indoor} (short range) and \textit{outdoor} (longer range) experimental setup (Figure~\ref{fig:indoor_outdoor_setup}). Both setups consider a line-of-sight (LOS) scenario with only two devices, an \textit{initiator} device, and a \textit{responder} device. As per Figure~\ref{fig:setup_distances}, in the indoor experimental setup the devices were placed on the ground and separated at distances of 0.5, 1.0, and 1.5 meters, while in the outdoor experimental setup, the devices were placed on tripods in an open area and separated at 3.0, 5.0, and 10.0 meters.

\begin{figure*}[t]
\centering  
    \begin{subfigure}[t]{1.0\columnwidth}
        \centering
        \includegraphics[width=1.0\columnwidth]{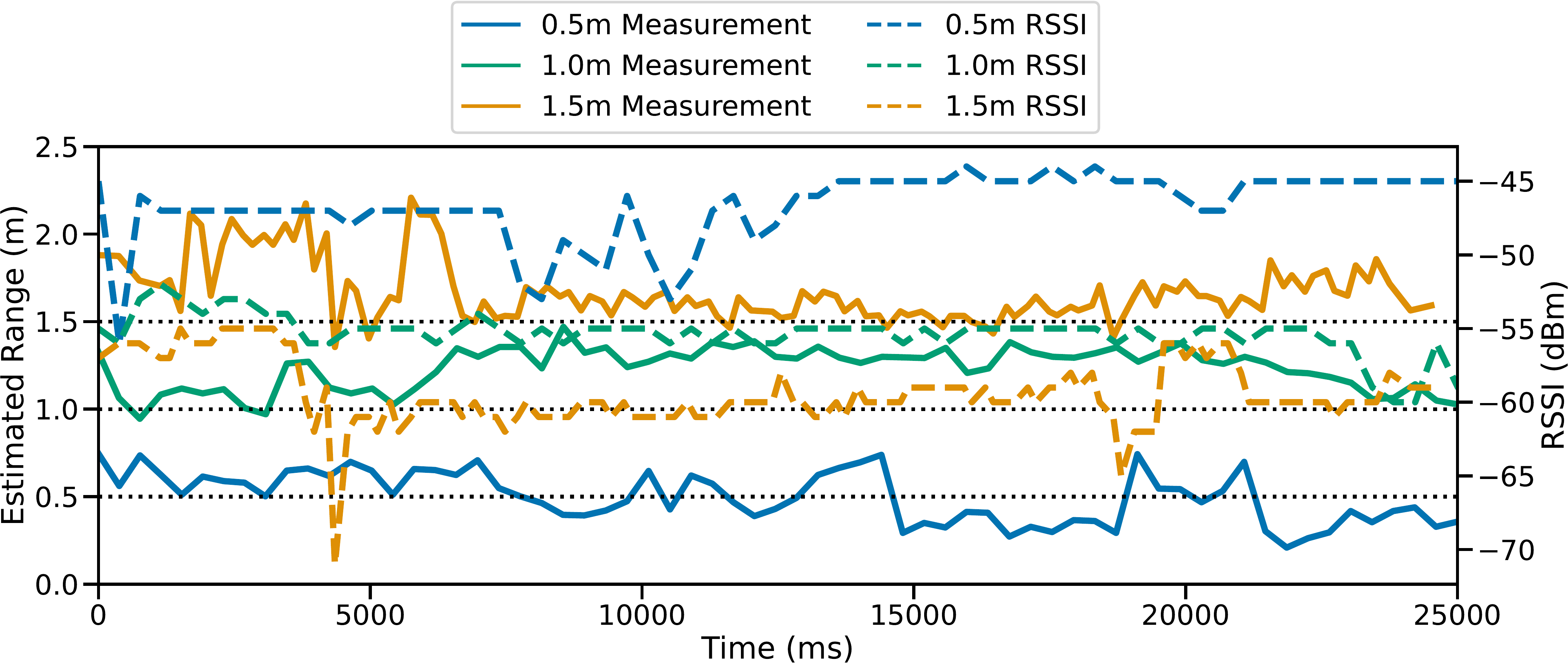}
        \caption{Estimated range and RSSI over time (indoors).}
        \vspace{4.00mm}
        \label{fig:timeseries_indoor}
    \end{subfigure}\hfill
    \begin{subfigure}[t]{1.0\columnwidth}
        \centering
        \includegraphics[width=1.0\columnwidth]{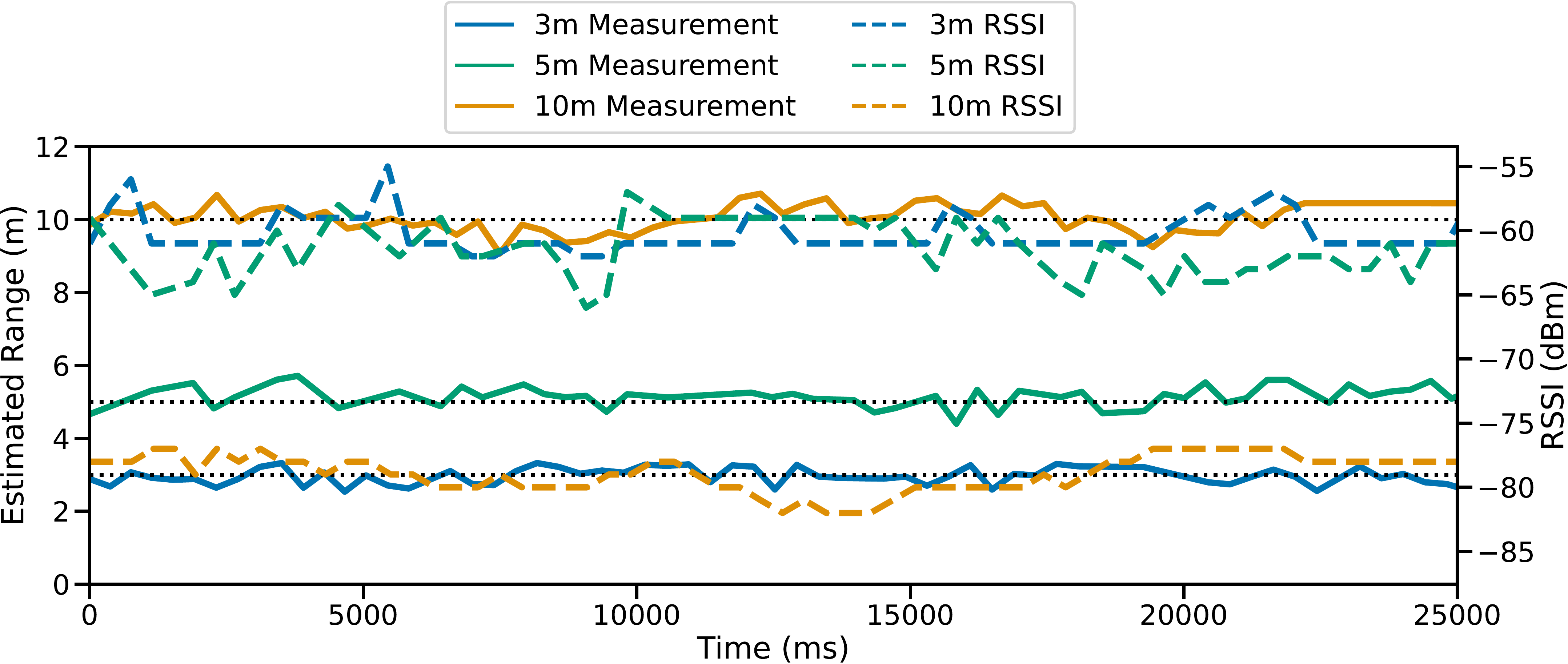}
        \caption{Estimated range and RSSI over time (outdoors).}
        \vspace{4.00mm}
        \label{fig:timeseries_outdoor}
    \end{subfigure}
    \begin{subfigure}[t]{1.0\columnwidth}
        \centering
	\includegraphics[width=1.0\columnwidth]{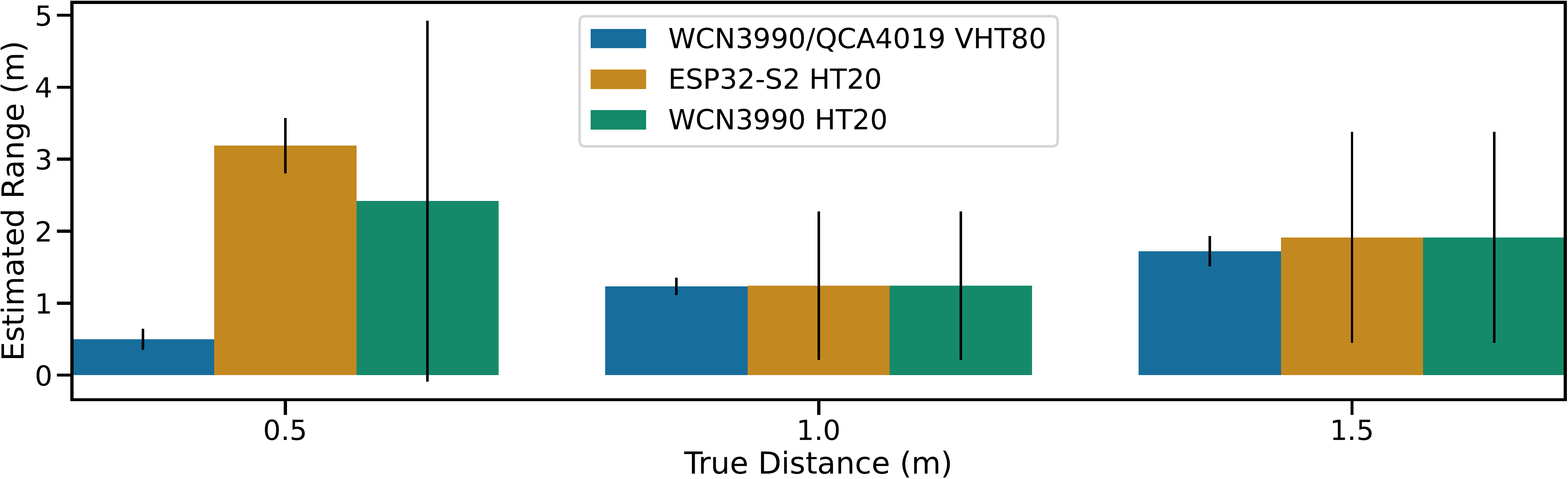}
	\caption{Mean \ftm estimated range (indoors).}
	\label{fig:measurements_indoor}
    \end{subfigure}\hfill
    \begin{subfigure}[t]{1.0\columnwidth}
        \centering
	\includegraphics[width=1.0\columnwidth]{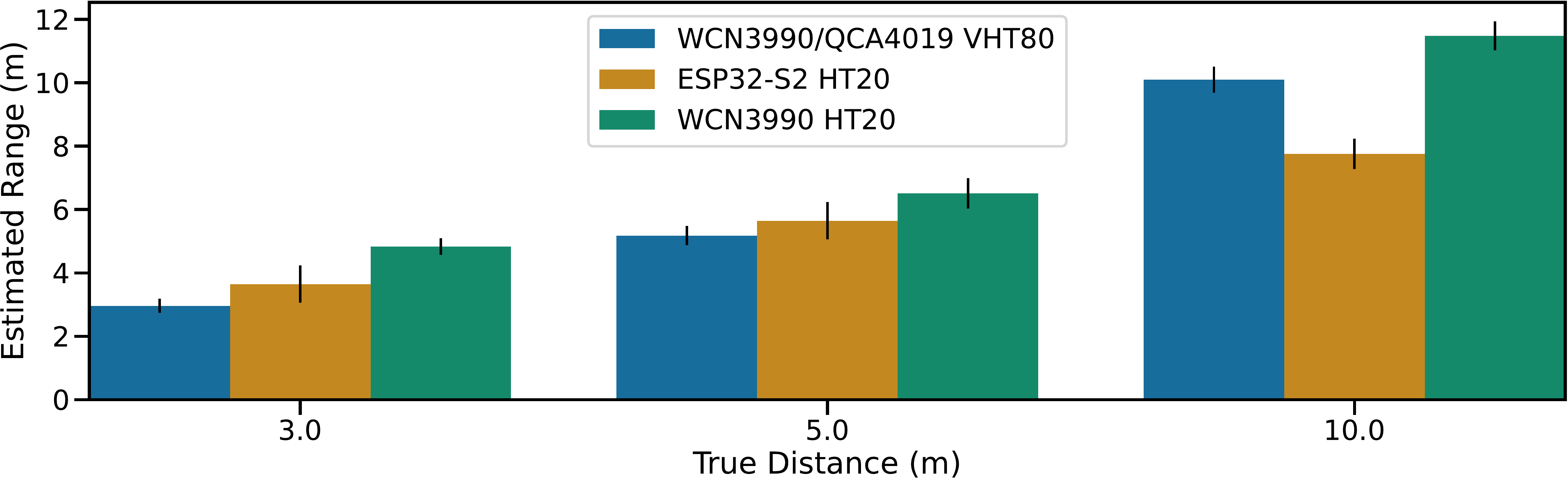}
	\caption{Mean \ftm estimated range (outdoors).}
	\label{fig:measurements_outdoor}
    \end{subfigure}
    \caption{\ftm ranging results for both indoor and outdoor experimental setups. Samples were taken every 380ms over a period of 25s. Using VHT80 allows accurate ranging even at short distances, while HT20 chipsets struggle at such close range. }
    \vspace{-4.00mm}
    \label{fig:results_indoor_outdoor}
\end{figure*}

Only the \texttt{ESP32-S2} and \texttt{WCN3990} support both $I$ and $R$ modes allowing us to use the same device for the initiator and responder; when using `Native' \ftm ranging on the \texttt{ESP32-S2} and using \aware on the \texttt{WCN3990}. Unfortunately, in this configuration, these devices are limited to 2.4GHz and HT20 (20MHz channel bandwidth). However, by employing a Google \wifi Mesh AP (\texttt{QCA4019}) for the \textit{responder} and the \texttt{WCN3990} for the \textit{initiator} (allowing one to use `Native' mode rather than \aware), it is possible to perform \ftm ranging at 5GHz on VHT80 (80MHz channel bandwidth). These configurations are summarized in Table~\ref{tab:setup_configurations}, and all experimental results are presented in Figure~\ref{fig:results_indoor_outdoor}.

%---%
\boldpar{Config 1 (VHT80 + `native' ranging)} The first configuration uses a Google Pixel 4a (\texttt{WCN3990}) as \textit{initiator} STA and a Google Wi-Fi MESH AP (\texttt{QCA4019}) as \textit{responder} the responder AP. The \wifi baseband in both chipsets supports IEEE\,802.11ac with VHT80, allowing testing of \ftm ranging over a wider (80MHz) channel bandwidth. From Figure~\ref{fig:results_indoor_outdoor} it can be seen that for both the indoor and outdoor setups the \texttt{WCN3990} manages to estimate the distance to the responder with sub-meter accuracy at all distances, and a small standard deviation.

% Despite the fact that this level of measurement error or noise contributes to the overall error, we calculated the median of numerous observations and found that the standard deviation of the FTM RTT data is much smaller than it is in the range measurements as shown in Figure~\ref{fig:results_indoor_outdoor}. 
% In the first configuration, the measurement error or noise contributes very little to the overall inaccuracy in VHT80 mode FTM ranging.

%---%
\boldpar{Config 2 (HT20 + `native' ranging)} The second configuration employs two FeatherS2 NEO (\texttt{ESP32-S2}) boards as both \textit{initiator} (STA) and \textit{responder} (AP) devices. The \texttt{ESP32-S2} can be clocked at up to 240 MHz in 2.4 GHz HT40 mode, however \ftm ranging is only supported on HT20\footnote{https://github.com/espressif/esp-idf/tree/master/examples/wifi/ftm}. While examination of the results in Figure~\ref{fig:measurements_outdoor} shows that the \texttt{ESP32-S2} is fairly accurate in the outdoor scenarios at longer ranges, at short distances indoors (Figure~\ref{fig:measurements_indoor}) it performs exceptionally poorly -- with an error of $\approx$1m at a distance of 0.5m, and an error of 2-3m at a true distance of 1.0m and 1.5m. While the indoor environment suffers from external interference sources and multipath reflects, the \texttt{ESP32-S2} suffers from poor receiver sensitivity short distances\footnote{https://tinyurl.com/yc2ca746}. It is therefore likely that the main factors which might be contributing to inaccuracy are lower bandwidth (i.e. HT20 mode), first arrival correction errors, and hardware timestamping delays in FTM packets.

%---%
\boldpar{Config 3 (HT20 + \aware ranging)} The final configuration considers \aware-supported FTM ranging~\cite{singh2022reliable} using one NAN anchor master and one NAN slave device using two Google Pixel 4a devices (\texttt{WCN3980}). As with The \texttt{ESP32-S2}, the \texttt{WCN3980} is limited to HT20 mode as this is the only mode supported by \aware for \ftm ranging in the Android framework. Equally, the \texttt{WCN3980} performs well in the outdoor scenario at longer distances (Figure~\ref{fig:measurements_outdoor}) with measurements accurate to within a meter. However, at shorter distances (Figure~\ref{fig:measurements_indoor}) the ranging measurements become wholly inaccurate -- deviating from the true distance with errors up to multiple meters. Again, while there is external interference and reflections in the indoor environment, the culprit for this poor accuracy is likely the lower receiver sensitivity due to the use of HT20 mode in this configuration.

\section{Security Analysis}
\label{sec:security}
\noindent From a security perspective, Wi-Fi FTM can be useful for identifying and locating rogue devices that may be attempting to gain unauthorized access to a network. However, it also has the potential to be exploited by attackers to perform location tracking and other forms of surveillance on legitimate devices connected to the network. Specifically, \ftm protocol messaging is based on a request-response model, therefore relying on the integrity of the responding device.

By default, \ftm control frames are unprotected and no cryptography and authentication is applied on the top messaging protocol in unassociated ranging mode. \ftm can be used to \textbf{track the location} of devices on a network by measuring the time it takes for a signal to travel between a device and an access point. Attackers can exploit this to perform location tracking on a legitimate device. The FTM protocol is vulnerable to \textbf{replay attacks}, where an attacker can intercept and replay a valid FTM response to a requesting device in order to impersonate a legitimate device and gain access to the network. While \ftm can detect \textbf{rogue devices} on a network by measuring the timing of the signals they send and comparing them to the timing of signals from legitimate devices, attackers can also use this feature to bypass security measures. An attacker that can \textbf{impersonate a device} can use the FTM feature to get information about other devices in the network which can be used to perform further attacks. 

To mitigate these risks, it is important to ensure that \ftm ranging is used in the associated mode using WPA3 or WPA2+PMF to improve FTM packet privacy protection. Moreover, a Packet Number (PN) check at LMAC firmware is expected to prevent replay attacks. At the ranging application layer, multi-factor authentication is expected to mitigate issues related to the elevation of privilege. For unassociated ranging MAC address randomization is expected to provide some basic security measures. Table~\ref{tab:vulnerabilities} outlines the main \ftm vulnerabilities that we have identified in this paper, as well as possible mitigations.

\begin{table}[t]
    \caption{\ftm threats and possible mitigations.}
    \label{tab:vulnerabilities}
    \footnotesize
    \renewcommand{\arraystretch}{.9}
    \centering
    \begin{tabular}{ c l }
        \toprule
        \textbf{Threat}         & \textbf{Mitigation}  \\
        \midrule
        Location tracking       & WPA3 or WPA2+PMF (802.11w)        \\ 
        Replay attacks          & PN check (LMAC)                   \\ 
        Rogue Device Detection  & Associated ranging with rekeying  \\ 
        Elevation of privilege  & Multifactor authentication        \\
    \end{tabular}
    % \vspace{-2.00mm}
\end{table}

\begin{figure}[t]
        \centering
        \includegraphics[width=.9\columnwidth]{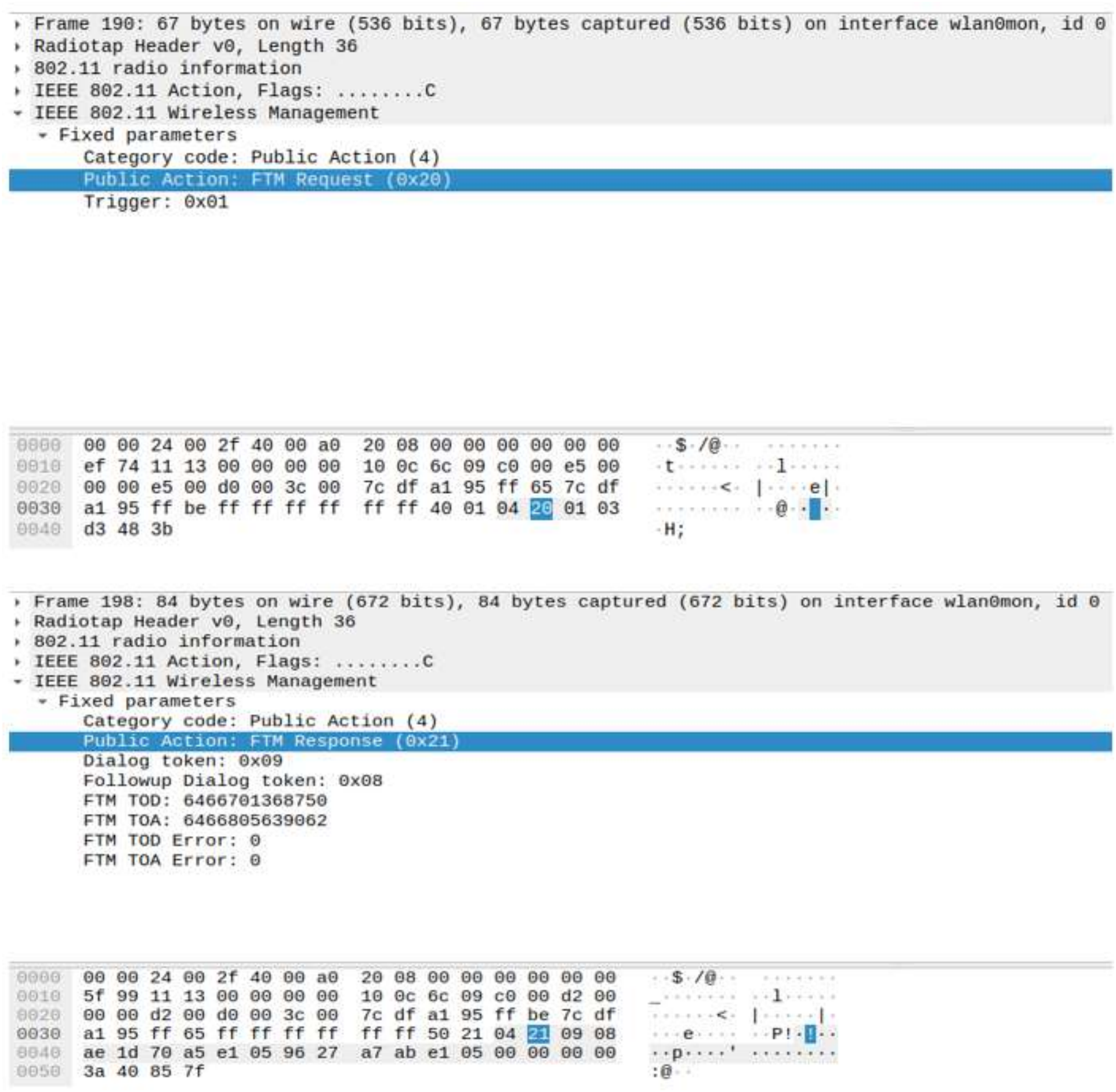}
        \caption{On-air \ftm ranging messages are unprotected and can be easily read by a nearby sniffer.}
        \vspace{-4.00mm}
        \label{fig:ftm_sniffer}
\end{figure}

\section{Related Work}
\label{sec:related_work}
\noindent WiFi FTM protocol ranging is getting widely adopted for many use-cases related to ranging, indoor localization for industrial robots, and asset tracking use cases in the industrial IoT segment. To date, there are many examples of commercial AP manufacturers enabling  FTM responder functionality to enable more precise asset tracking, time, and motion analysis~\cite{arubanetworks}. This includes several of the recent “mesh” APs (e.g., google Nest Wifi, Eero Pro, Netgear Orbi, Linksys Velop,  ASUS RT-ACRH13). Furthermore, FTM initiator functionality Google has worked with key wifi chip vendors(Qualcomm, Broadcom) to enable the complete AOSP ranging framework~\cite{Google} to integrate multilateration algorithm to estimate the absolute location in an indoor environment. Salomon~et~al.~\cite{salomon21csi} proposed the implementation of distance estimation approaches based on both RSSI and CSI measurements using the Nexmon CSI Extractor on Raspberry Pi 4 devices. In a  multi-radio environment,  the fusion-based scheme can be used to improve location in diverse ways to achieve higher precision with frequency band diversity~\cite{alvarez2021wifi}.

\section{Conclusions}
\label{sec:conclusions}
\noindent In this paper we have presented an overview of the parameters that can affect the accuracy of the FTM protocol and performed an experimental study using three different commercial \wifi chipsets supporting the \ftm protocol. We compared the ranging accuracy achieved in the different bandwidth configurations for a short distance (up to 1.5m) indoor scenario as well as a longer distance (up to 10m) outdoor scenario, finding that while the \ftm chipsets that support the wider channel bandwidths available in the 5GHz spectrum are capable of providing fairly accurate ranging estimations in both the indoor (short range) and outdoor (longer range) scenarios, \ftm ranging at 2.4GHz on the narrower HT20 bandwidths performs poorly in the indoor setting at shorter distances, where believe that the lower receiver sensitivity of the HT20 devices is a significant factor. While we were unable to comprehensively evaluate all of the \wifi cards which currently support FTM, these findings show that \ftm deployments should strongly consider using 5GHz VHT80 configurations for accuracy at both longer and shorter ranges. Furthermore, by sniffing the on-air packets, we have demonstrated that \ftm messaging is completely unprotected and could easily be tampered with. These findings suggest that as well as studying upcoming improvements to \ftm accuracy through the IEEE\,802.11az specification, which leverages MAC and PHY-level techniques, further work is needed to explore the \ftm security concerns and mitigations we have outlined in our threat model.

% use section* for acknowledgements
% \section*{Acknowledgements}

% Acknowledgements go here

% References section
% \bibliographystyle{IEEEtran}
% \vspace{2.00mm}
\bibliographystyle{unsrt}
\bibliography{references}

% that's all folks
\end{document}